# Biochemical Filter with Sigmoidal Response: Increasing the Complexity of Biomolecular Logic

Vladimir Privman, Jan Halámek, Mary A. Arugula,

Dmitriy Melnikov, Vera Bocharova and Evgeny Katz*

*Department of Chemistry and Biomolecular Science, and*

*Department of Physics, Clarkson University, Potsdam, NY 13699*

---

**\*Corresponding author:**

E-mail: ekatz@clarkson.edu; Tel.: +1 (315) 268-4421; Fax: +1 (315) 268-6610

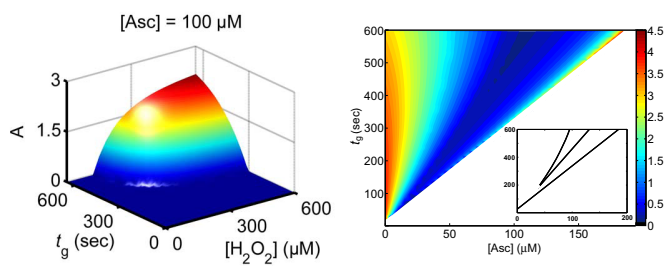

**ABSTRACT**

The first realization of a designed, rather than natural, biochemical filter process is reported and analyzed as a promising network component for increasing the complexity of biomolecular logic systems. Key challenge in biochemical logic research has been achieving scalability for complex network designs. Various logic gates have been realized, but a "toolbox" of analog elements for interconnectivity and signal processing has remained elusive. Filters are important as network elements that allow control of noise in signal transmission and conversion. We report a versatile biochemical filtering mechanism designed to have sigmoidal response in combination with signal-conversion process. Horseradish peroxidase-catalyzed oxidation of chromogenic electron donor by $H_2O_2$, was altered by adding ascorbate, allowing to selectively suppress the output signal, modifying the response from convex to sigmoidal. A kinetic model was developed for evaluation of the quality of filtering. The results offer improved capabilities for design of scalable biomolecular information processing systems.

## 1. Introduction

Biochemical processes,[1-8] and more generally, chemical kinetics,[9-15] have been actively researched for novel paradigms of information processing. The promise of biochemical computing has ranged from *in situ* decision making,[16,17] novel multi-input biosensors,[18-22] for instance, for signaling in cases of trauma/injury,[23-25] to bioelectronic devices[26-28] and actuators,[29] and ultimately to interfacing of living organisms with Si electronics. A key challenge has been increasing the *complexity* of biochemical computing systems while maintaining fault-tolerant, low-noise, scalable "network" functionality.[8,30] Nature, of course, offers a paradigm for complex information processing with biomolecules. However, realizable man-made networks of concatenated chemical and biochemical reactions, frequently based on enzyme-catalyzed processes, are presently far from the demands of "bottom-up" design of complex "artificial life" systems by mimicking natural processes. A potentially more practical approach has been to turn to the well-established scalability paradigm of Si electronics, aiming at digital information processing with binary-logic gates and their networks. There has been a substantial recent effort aimed at realizing gates such as AND, OR, XOR, etc., through relevant biochemical kinetics.[6,7,31-50] Few-gate networks,[27,51,52] as well as interfacing of enzyme-based biochemical logic with Si electronics,[28] and certain functional units for memory,[53] arithmetic operations,[54] and security and control devices,[55-57] have been demonstrated, as recently reviewed.[8,46]

Presently, biochemical information processing systems are not intended as a replacement of Si devices, but rather aim at offering additional functionalities in situations where direct wiring to computers and power sources is not practical such as in many biomedical applications.[23-25] However, even for near-term applications, scalable and versatile networking paradigms are crucial. Recent studies suggest[8,58,59] that the level of noise in biochemical systems is quite high as compared to electronics. This includes noise both the input/output signals and in the "gate machinery" chemical (e.g., enzyme) concentrations. Avoiding noise amplification by appropriate network design is therefore quite important even for small networks, similar to recent findings[60] for networking of neurons. Present estimates[8,61] suggest that, not only analog but also digital error correction will be required for networks involving more than order 10 processing steps.

Considerations of scalability and control of noise have set the stage for new challenges. Large-scale interconnectivity and fault-tolerance cannot be achieved without the development of a "toolbox" of new network elements including filters, signal splitters, signal balancers, resetting functions, etc. These analog network elements for biochemical computing might not follow too closely the device components of Si electronics. In fact, concepts borrowed from natural systems, specifically, memory involving processes[62] have recently received attention in unconventional information processing studies. However, as a rule none of the standard elements for networking for (ultimately, digital) information processing has been experimentally realized to date in a setting demonstrating interconnectivity with binary logic gates.

In this work, we report the first experimental realization of a biochemical filter, as well as its modeling within a kinetic description of the (bio)chemical reactions involved. Filtering involves passing the signal through a network element with a sigmoidal response curve. The analog input values, spread about the reference 0 and 1, are thus pushed closer towards the 0 and 1 of the output, respectively; see Fig. 1. Such functions, as components utilized in combination with other tools, for instance, for signal splitting and redundancy, are crucial for fault-tolerant network design in the analog-digital information processing paradigm contemplated for biochemical gate-based logic. Thus, it is important not only to devise and experimentally realize sigmoidal-response filters, but also to accomplish this in settings which demonstrate potential interconnectivity with logic gates. Model analysis of our experiment helps identify why it has been so challenging to realize "man made" biochemical filtering systems. Indeed, the (bio)chemical processes utilized are standard, which is actually advantageous for versatility and in applications. However, we find that the low noise-scaling-factor region of high-quality filtering is realized close to the large-intrinsic-noise regime (small signal range). A careful selection of process parameters, facilitated by modeling is thus required for realizing the filtering effect.

## 2. Experimental

***Chemicals and Reagents.*** Peroxidase from horseradish type VI (HRP, E.C. 1.11.1.7), hydrogen peroxide 30% wt ACS reagent ($H_2O_2$), L-ascorbic acid, and 3,3',5,5'-tetramethylbenzidine (TMB) were purchased from Sigma-Aldrich and used as supplied. Ultrapure deionized water (18.2 MΩ·cm) from a NANOpure Diamond (Barnstead) source was used in all our experiments.

***Signal Definition and Measurement.*** We took $H_2O_2$ as the logic input. The chemical reaction was catalyzed by HRP and filtering effect was accomplished with the added ascorbate (Asc). The logic output was measured as the concentration of the charge transfer complex of TMB and $TMB_{ox}$, which was detected by measuring the absorbance, $A$. The absorbance measurements were performed using UV-2401PC/2501PC UV-visible spectrophotometer (Shimadzu, Tokyo, Japan) at 37°C. All measurements were performed in 1 mL poly(methyl methacrylate) (PMMA) cuvettes in 0.01 M phosphate buffer, pH = 7.69. TMB, 1 mg/mL, was dissolved in DMSO and added to the reaction solution and thoroughly mixed using a pipette. The production of TMB charge transfer complex as a function of the reaction time was measured at wavelength $\lambda = 655\,\text{nm}$. The absorbance values were converted to concentrations using extinction coefficient $\varepsilon_{655} = 39\,(\text{mM}\cdot\text{cm})^{-1}$.

## 3. Results and Discussion

***Statement of the Problem.*** In order to facilitate the discussion and identify potential challenges in realizing the desired systems, let us consider a biochemical reaction with enzyme, *E*, as the biocatalyst. As the simplest model, we will for now assume specific, irreversible reaction steps with one intermediate compound, *C*,

$$E + I \xrightarrow{R} C$$
$$C + S \xrightarrow{r} E + P \qquad (1)$$

where $I$ is the substrate the initial concentration of which, $I(0)$, will be regarded at the input signal, and, for closer correspondence with our experiment, we assume the intake of the second substrate, $S$. Here $P$ denotes the final product, whereas $R$ and $r$ are rates constants. Of course, the actual reaction pathways for most enzymes, including horseradish peroxidase (HRP) used in our experimental study, are more complicated and not all are irreversible. We consider the concentration of the product, $P(t_g)$, at a specific "gate" time, $t_g > 0$, as the output. Then the enzymatic reaction mimics the simplest possible gate function: the identity, i.e., signal transmission or conversion. Indeed, there is no output signal for the initial concentrations of the input at zero, whereas the output reaches concentration $P_{max}(t_g)$ when the input is supplied at $I_{max}(0) > 0$. Thus, we take zero concentrations as logic-0, and the reference input logic-1 value marked with "max". In biomedical applications, the reference input-1 concentrations are usually fixed at the average values corresponding to, for example, elevated pathophysiological conditions,[25] and, in fact, the input-0 concentrations can be at normal values rather than at the physical zero. The corresponding output "logic" values are set by the gate function itself.

In terms of the rescaled variables that define the range between the logic 0 and 1, here $x = I(0)/I_{max}(0)$, $z = P(t_g)/P_{max}(t_g)$, we consider the response-curve function, $z(x)$, which connects the logic-point values. Note that in many AND-gate realizations, enzymatic reactions were used with both substrates as varied inputs, and with the two-argument response-surface $z(x, y)$. In applications with the "logic" intervals not starting at the physical zeros, subtractions are needed to define $x$, $y$, $z$. Furthermore, ranges rather than sharp values have to be considered, likely somewhat different for the same logic output at different combinations of inputs (for example, the OR gate has three outputs at logic-1). All this can be viewed as part of the various sources of random and built-in "intrinsic" noise that should be filtered out for the ultimate binary decision-making for "field" diagnosis of the "action/no-action" alert type, involving multi-input sensor applications.[1,8,23-25] Such noise would be more straightforward to handle in electronic systems, but in biochemical systems it is orders of magnitude larger and much more difficult to suppress even in very small networks, well below the size that would necessitate digital error correction based on redundancy.

The shape of the response curve (surface), notably its slope near the logic points, if larger than 1, is important in network-element functioning (because of analog noise amplification). Since these shapes are convex for most biocatalytic reactions, they typically amplify noise at least at one logic point, as network elements. This is shown in Fig. 1 for the model of Eq. 1, whereas details of our experimental system will be introduced later. The gate-response curve/surface is typically convex because in enzymatic processes, for small inputs, $I(0)$, the output, $P(t_\mathrm{g})$, is usually proportional to the input, whereas for large inputs the output signal reaches saturation by exhausting the activity of the biocatalyst and due to limited availability of other chemicals, including $S(0)$. Optimization by attempting to change the shape of the response curves/surfaces without modifying the system, has proved difficult because in terms of the rescaled variables, such as $x$ and $z$, the main, linear effects of varying the reaction rates, $R$, $r$, and the initial enzyme concentration, etc., are largely cancelled out. The higher-order "nonlinear" effects, require substantial changes, by orders of magnitude, which are in most situations not experimentally feasible. Consideration of enzymatic systems functioning as AND gates with non-smooth response surfaces or those with sigmoidal response due to self-promoting property of one of the inputs (applicable for instance for many allosteric enzymes), has been reported but did not yield a variety of systems beyond few isolated examples of gates.[8,58] Thus, we are faced with the need to develop simple, versatile kinetic mechanisms for converting convex response curves, such as the one illustrated in Fig. 1, to sigmoidal, by actually modifying the set of the chemicals and their reactions by, here, adding another process to the "gate" system. This can be viewed as incorporation of another "network element" in our chemical-soup biocatalytic system of reactions.

***Biochemical Filter.*** Since typical biocatalytic reactions always reach saturation (flat region) for a range of large inputs, the primary challenge has been to design and experimentally realize a generic reaction scheme the kinetics of which can eliminate the linear buildup of the signal at small inputs. A promising mechanism[61] apparently used by Nature,[63] can involve the introduction of an additional chemical, *F*. This filter-effect-causing reactant should actively neutralize a fraction of the output. However, its initial quantity, $F(0)$, should be quite limited so that, by the gate time $t_g$, its supply is exhausted except for relatively small values of the input.

Schematically, we add to the processes in Eq. 1, the reaction $P + F \xrightarrow{\rho} \ldots$, with the products that could be inert chemicals, and the rate constant of which, $\rho$, is relatively large. While theoretically feasible, this approach poses a problem that it then generally weakens the output signal. To compensate, in order to preserve the large-input-side saturation property, the input reference "logic-1" value must be nontrivially increased. This, however, might not be feasible in applications. Furthermore, too much input reactant might cause undesirable (bio)chemical effects. For example, in our system a sufficiently large quantity of the input, hydrogen peroxide, can actually inhibit the activity of HRP.

The products of the added reaction could also include chemicals which are active in other parts of the system. For example, reactions of the type $P + F \xrightarrow{\rho} I + \ldots$ introduce a *feedback loop*. Here we consider another variant,

$$P + F \xrightarrow{\rho} S + \ldots \tag{2}$$

with one of the produced chemicals being the second substrate. Indeed, reversing the last step of the chemical transformation leading to the product *P*, seems to be the easiest added process to practically realize, and the least disruptive for the other chemical kinetics steps. It also offers the back-supply of one of the initial reactants, thus to an extent compensating for a possible reduction in the overall output strength. In our case, the second substrate, 3,3',5,5'-tetramethylbenzidine (TMB) serves as the chromogen the oxidation of which ultimately results in optically detectable compounds. The added process then involves the reduction caused by the introduced ascorbate as an ionic species constituting the reactant *F*. Fig. 1 illustrates that the system of Eqs. 1-2 can indeed lead to sigmoidal behavior.

Before turning to the actual experiment and its modeling, let us point out another important requirement for adding steps to the system. The new reagent, $F$, should ideally not only *produce* active chemicals selectively, here $S$ in Eq. 2, but it should also *not react* with various other chemicals but the intended one: the product, $P$. Generally cross-talk of (bio)chemical reactions is an important issue in biocomputing design which ultimately, for large networks, will require spatial separation by compartmentalizing the process steps in microfluidic systems, at electrodes and in layers of their structure, etc.[46,64,65] Generally, some limited degree of reversibility might be unavoidable in reactions such as Eq. 2. In our experimental system the cross-talk in negligible and the added reaction is irreversible.

***Experimental Data.*** Our experimental system has been alluded to in Fig. 1. Enzyme HRP consumes the input-signal, $H_2O_2$, and oxidizes the chromogen, TMB, with and without the filter-effect-causing reactant added, ascorbate (Asc). The output signal is detected optically as a blue charge-transfer complex of TMB and $TMB_{ox}$,[66] by measuring the absorbance, $A$, as shown in Fig. 2. The use of TMB (in the presence of HRP) as a detector for $H_2O_2$ is quite common because of its extreme sensitivity to even the slightest amounts of $H_2O_2$, and also because it is clinically safe. The HRP-catalyzed oxidation of TMB is a complicated, multistage process with two colored products: the yellow-colored oxidized form $TMB_{ox}$, and the blue-colored charge transfer complex $TMB \cdot TMB_{ox}$.[66] The latter exists in equilibrium with $TMB_{ox}$ and radical cations $TMB^{+\cdot}$. The radicals are produced in the normal peroxidase cycle.[66,67] We point out that the mechanism of action of HRP is rather complicated[68,69] and involves more than a single intermediate complex, with the associated reaction pathways. Our parameter selection for the system was to a large extent based on experience, on the experimental convenience, and on the regimes quoted in the literature. However, we had to avoid taking too much $H_2O_2$ to prevent inhibition of HRP.

Response-curve measurements are presented in Fig. 3, as data for four respective sets of experiments, with 0, 60, 100, and 200 µM of Asc introduced initially. The initial concentration of $H_2O_2$ (logic input) was varied from 0 (or somewhat above that value when [Asc](0) was high) to 600 µM (and 10% above that value when [Asc](0) was 0), while the initial concentrations of TMB, 0.42 mM, and HRP, 0.44 nM (5 U/L), were kept constant. The data represent time-dependence measurements, with [H$_2$O$_2$](0) covering the range of its values in equal steps: there are total 14, 13, 12, 11 time-dependent data set in the plots in Fig. 3, in order of the increasing initial [Asc], respectively.

***The "Identity" Gate Function.*** While measurements of the response curves are needed for our study of the filtering functionality and noise handling, it is useful to also further comment on the signal selection in connection with the original intent of the system as the "identity" gate. The "digital" signal is observed if the input, H$_2$O$_2$, is initially present in the solution at a pre-selected "logic-1" concentration value, and is not observed otherwise. Some examples are given in Fig. 4. Note that the separation between the logic-0 and logic-1 for the two examples shown (with and without the ascorbate added), is the largest at 655 nm (see also Fig. 2).

***Kinetic Modeling.*** In the selected regime of parameters, the following kinetic model can be used to keep the number of adjustable rate constants manageable. We consider the main reaction steps, were the minuses refer to the back-reaction rates, and, despite the prefactors 2 in some places for clarity, these are not fully species- and charge-balanced chemical process schemes, but rather schematics:

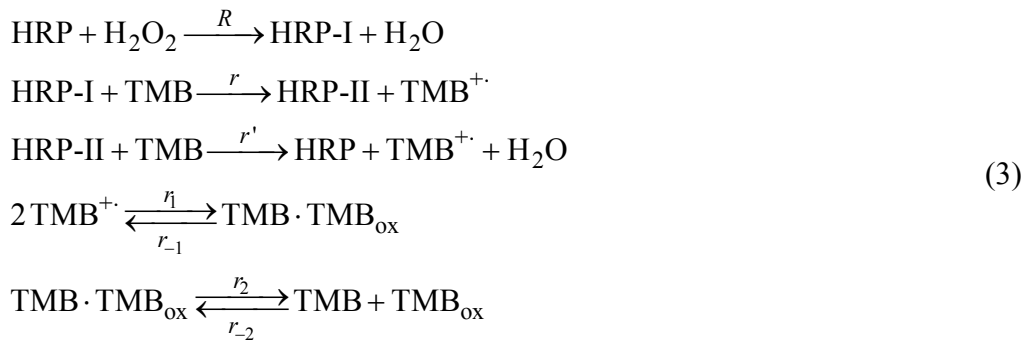

$$\begin{aligned} \text{HRP} + \text{H}_2\text{O}_2 &\xrightarrow{R} \text{HRP-I} + \text{H}_2\text{O} \\ \text{HRP-I} + \text{TMB} &\xrightarrow{r} \text{HRP-II} + \text{TMB}^{+\cdot} \\ \text{HRP-II} + \text{TMB} &\xrightarrow{r'} \text{HRP} + \text{TMB}^{+\cdot} + \text{H}_2\text{O} \\ 2\,\text{TMB}^{+\cdot} &\underset{r_{-1}}{\overset{r_1}{\rightleftarrows}} \text{TMB} \cdot \text{TMB}_{\text{ox}} \\ \text{TMB} \cdot \text{TMB}_{\text{ox}} &\underset{r_{-2}}{\overset{r_2}{\rightleftarrows}} \text{TMB} + \text{TMB}_{\text{ox}} \end{aligned} \qquad (3)$$

The first three reactions describe the peroxidase cycle with the radical cation as the output. We assume irreversible steps, and both intermediate products, HRP-I and HRP-II will be replaced by a single effective complex, earlier introduced as $C(t)$ in Eq. 1. Indeed, here HRP-I is produced practically irreversibly and also fast as compared to its rate of conversion into the second intermediate complex, HRP-II, as is known for a similar peroxidase system.[67] Furthermore, the process with rate constant $r'$ is expected to be significantly slower than that with $r$.[67] Therefore, we can use only one effective rate constant, $r$, as in Eq. 1, because this process "drives" the next process, of rate constant $r'$ (with the counting of the produced radical molecules properly doubled). The last two processes correspond to the formation and interconversion of the colored products. These reactions are reversible, but for our modeling, the available data were not detailed enough to fit so many parameters, and we thus set $r_{-1}, r_{-2} = 0$, which is justified by the experimentally demonstrated fact that the back-reactions here are slow.[66] This simplification does not correctly describe the large-time limit, but we do not reach this limit in our experiments (as can be seen in Fig. 2). Reaction times here, $t < 600$ sec, also favor the use of the blue charge-transfer compound (which appears first) to define the output signal, instead of the yellow $TMB_{ox}$, which appears later in the reaction. Finally, the sigmoidal property is induced by adding ascorbate that irreversibly converts the blue compound back into TMB,

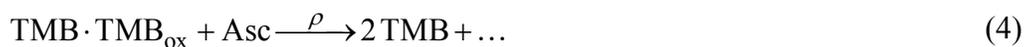

$$\text{TMB} \cdot \text{TMB}_{\text{ox}} + \text{Asc} \xrightarrow{\rho} 2\,\text{TMB} + \ldots \qquad (4)$$

which is a fast reaction. Ascorbate also reacts with $TMB_{ox}$, but we ignore this process for our regime of relatively small reaction times.

The actual rate equations used in data fitting, detail the schematics, just discussed in connection to Eqs. 3-4, of the assumed reaction steps,

$$\frac{dI(t)}{dt} = -RI(t)[E(0) - C(t)]$$

$$\frac{dC(t)}{dt} = RI(t)[E(0) - C(t)] - rS(t)C(t)$$

$$\frac{dS(t)}{dt} = -2rS(t)C(t) + r_2[\text{TMB} \cdot \text{TMB}_{\text{ox}}](t)$$
$$\qquad + 2\rho F(t)[\text{TMB} \cdot \text{TMB}_{\text{ox}}](t)$$

$$\frac{d[\text{TMB} \cdot \text{TMB}_{\text{ox}}](t)}{dt} = r_1 \{[\text{TMB}^{+\cdot}](t)\}^2 - r_2[\text{TMB} \cdot \text{TMB}_{\text{ox}}](t) \qquad (5)$$
$$\qquad - \rho F(t)[\text{TMB} \cdot \text{TMB}_{\text{ox}}](t)$$

$$\frac{d[\text{TMB}^{+\cdot}](t)}{dt} = 2rS(t)C(t) - 2r_1 \{[\text{TMB}^{+\cdot}](t)\}^2$$

$$\frac{d[\text{TMB}_{\text{ox}}](t)}{dt} = r_2[\text{TMB} \cdot \text{TMB}_{\text{ox}}](t)$$

$$\frac{dF(t)}{dt} = -\rho F(t)[\text{TMB} \cdot \text{TMB}_{\text{ox}}](t)$$

Here the earlier introduced notation is utilized, with $I(t)$ denoting the concentration of $H_2O_2$, $S(t)$ denoting the concentration of TMB, $F(t)$ — that of ascorbate, and $C(t)$ — of the HRP-I complex. The initial concentration of the enzyme, $E(0)$, as well as those of $H_2O_2$, TMB and ascorbate are all known for each experimental time-dependence data set taken (e.g., Fig. 3). Time-dependent concentrations of all the other chemicals in the rate equations, initially (at time $t = 0$) are zero.

*Data Fitting.* Fig. 3 illustrates our main result: the emergence of the sigmoidal behavior as measured experimentally for various process times. For a range of times, each constant-time, $t = t_g$, slice in the plots in Fig. 3 displays, up to noise in the data, a convex (without ascorbate) or sigmoidal (when ascorbate is added) response curve. The data were fitted according to Eqs. 3-5, and the resulting plots are shown in Fig. 5. The measured quantity reflects the dependence of the charge-transfer complex concentration on the initial hydrogen peroxide concentration and reaction time, mapped out for four different initial ascorbate concentrations: $F(0) = [\text{Asc}](0) = 0, 60, 100$ and $200$ μM (Fig. 3).

While our focus here is on the input-signal, $I(0) = [H_2O_2](0)$ dependence, let us also comment on the time dependence: Specifically, note that in the plot with no ascorbate (see Fig. 3), one can discern that $[TMB \cdot TMB_{ox}](t)$ reaches a flat maximum at about 300 sec, and then actually somewhat decreases. This occurs because at larger times conversion of the measured blue product into yellow becomes nonnegligible. The fixed-$t$ response curves in this case (no ascorbate) have a convex shape, as expected for a standard enzymatic reaction, here with the smaller slope at the logic-1 point. Addition of the filtering agent delays the effective onset of the output signal, as initially all the charge transfer complex is converted back into TMB. Indeed, from data fitting (of all the data, not just those without ascorbate added), we were able to determine the rate constants introduced in connection with the processes in Eq. 3: $R = 9.76\ (\mu M \cdot sec)^{-1}$, $r = 2.08\ (\mu M \cdot sec)^{-1}$, $r_1 = 1.00\ (\mu M \cdot sec)^{-1}$, $r_2 = 0.03\ sec^{-1}$, where the values for $R$ and $r$ are consistent with published data for a related system.[67] However, the rate constant in Eq. 4 is so large that the data fitting was not sensitive to its value as long as it is taken $\rho > 10\ (\mu M \cdot sec)^{-1}$. The latter value was used in drawing Fig. 5.

The fact that the rate $\rho$ in Eq. 4 is large, indicates that the slope of the fixed-time sigmoidal cross-sections in Figs. 3 and 5 is very small at the origin: the curves are practically flat as functions of the supplied $H_2O_2$, until there is enough input so that all the ascorbate is consumed. Then the output concentration begins to increase, and the response curve eventually, for larger inputs, is similar to the one measured without ascorbate, but shifted; see Figs. 3 and 5. This indicates that there is a trade-off in the quality of the "filtering" of noise: the shift of the response curve to lager $[H_2O_2](0)$ results in the slope at the logic-1 point (at the fixed $[H_2O_2](0)$ value identified as the reference logic-1 input) gradually increasing. Thus, while the steepness of the central inflection region does not change significantly, we do have to select a proper range of values for $[Asc](0)$ in order to have it centrally positioned for a balanced filtering effect near both "logic" values, 0 and 1. This approximate $[Asc](0)$ value to use will depend on the gate time, $t_g$, and also to some extent on the level of noise expected.

***Optimization of the Sigmoidal Gate Function.*** Let us assume for simplicity, a Gaussian input-signal distribution due to noise (but at $x = I(0)/I_{max}(0) = 0$ — half-Gaussian, one-sided), of spread, $\sigma_{in}$, taken the same for signals centered at 0 and 1. By using our fitted model parameters, we can then calculate the respective spreads of the $z = P(t_g)/P_{max}(t_g)$ output, $\sigma_{out}^{(0,1)} = [\langle z^2 \rangle - \langle z \rangle^2]^{1/2}$. Here the averages $\langle \cdots \rangle$ are over the distribution at 0 or 1 (denoted by superscripts in parentheses). Fig. 6 plots the larger of the two noise-scaling ratios, $\sigma_{out}^{max}/\sigma_{in} = \max_{i=0,1}[\sigma_{out}^{(i)}/\sigma_{in}]$, for $\sigma_{in} = 0.1$ as an illustrative case. Note that in the lower-right part in Fig. 6, the amount of the blue output product is very small even at logic-1. In this regime the system is not useful as a "logic gate," because the noise in the curve $z(x)$ *values* (rather than the noise due to spread of the input signal about the reference 0 and 1) will be large on a relative scale. For the studied region of the reaction (gate) times and ascorbate concentrations that correspond to good resolution between the 0 and 1 values (the upper-left part in Fig. 6), the noise scaling factor varies from $\sim 3$ to $\sim 0.2$. Obviously we favor systems with significant noise suppression, $\sigma_{out}^{max}/\sigma_{in} < 1$, in the region of minimal scaling factor values, identified in the inset in Fig. 6. For example, for the gate time of 600 sec, the optimal amount of ascorbate is $\sim 120\ \mu M$. This range of values weakly depends on the input noise. Furthermore, for $\sigma_{in} = 0.3$ for instance, the noise scaling factor varies from $\sim 2$ to $\sim 0.7$, and, in fact, for larger noise, the filter effect will be lost. Our results suggest that the minimal values of $\sigma_{out}^{max}/\sigma_{in}$ exceed 1 beyond $\sigma_{in} \sim 0.4$.

Figure 6 also indicates that the optimal input ascorbate concentration depends on the reaction time: The smaller is the time, the less ascorbate is needed to minimize the spread of the analog noise in the output. This can also be discerned from the location of the inflection regions in the fixed-time slices in Figs. 3 and 5. While smaller amounts of ascorbate in the system can thus be advantageous, one should note that the physical separation between the logic 0 and 1 values also decreases at smaller times which means that the system becomes more susceptible to intrinsically generated noise. In fact, the low noise-scaling-factor region is always quite close to the large-intrinsic-noise regime (see the inset in Fig. 6), especially for short gate-times. This has been the main reason for the substantial challenge involved in demonstrating biochemical filtering. Indeed, as mentioned in the Introduction the idea is simple and the (bio)chemical processes utilized are standard, which is actually an advantage for the versatility of the identified

– 13 –

filter mechanism. However, a careful selection of parameter values, facilitated by modeling, seems to be crucial for experimentally realizing the filtering effect.

## 4. Conclusion

With Si electronics approaching its limits,[70] research efforts have turned not only to advanced nanostructures[71] and nanofeatures,[72] but also to alternative computing systems.[71,73-76] The latter are sought for speed-up of specific computations, for new information processing ideas, and for enabling capabilities. As mentioned in the Introduction, scaling up the *complexity* of information processing has been the primary challenge for most "unconventional" computing approaches being developed.

Here we considered information processing based on biochemical reactions, which not only offers a long-term futuristic promise of direct living-organism to Si-computer interface and perhaps variants of "artificial life," but also a shorter-term approach to improve multi-input, complex-decision-making biosensors[46] in field diagnostic biomedical applications.[23-25] We experimentally demonstrated, as well as identified by modeling considerations why is has been so challenging to accomplish, a key element for any "toolbox" for noise-tolerant networking for information processing: We realized a versatile *biochemical filter*. It is hoped that the results reported here will facilitate the development of the next-generation biomolecular logic systems of higher complexity, based not only on simple gates and their few-step concatenations, but also utilizing network element designs and other ideas form Si electronics, as well as from Nature. We anticipate that the concept of a "toolbox" of versatile functionalities such as the realized filter, will be instrumental in pushing the boundaries of biochemical/biomolecular computing as a viable, center-stage paradigm of unconventional information processing.

**Acknowledgements.** This work was supported by the NSF (CCF-1015983) and ONR (N00014-08-1-1202).

**FIGURES**

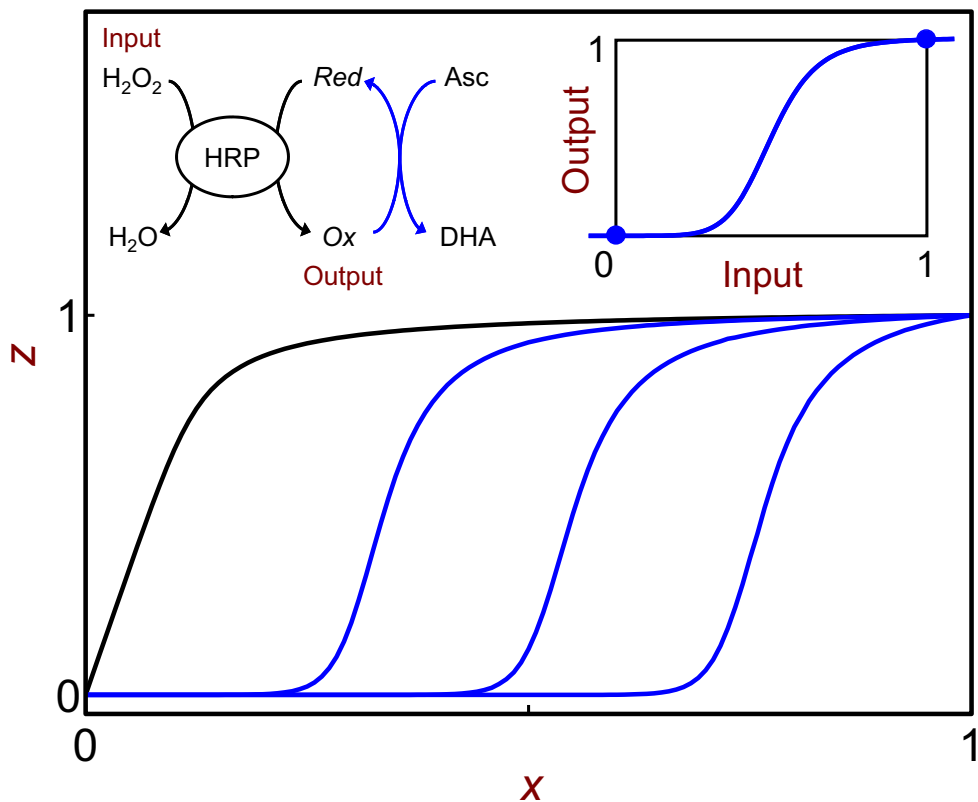

**Figure 1.** The convex and sigmoidal response for the "identity" logic gate mapping 0 to 0, and 1 to 1. The curves in the plot correspond to the model of Eq. 1: the top, convex curve, and to the same model with the added process, Eq. 2: the three sigmoidal curves with, from left to right, increasing $F(0)$. (Various other parameters in Eqs. 1-2 were conveniently selected for the illustration and are of no particular interest.) *The inset* illustrates an "ideal" sigmoidal curve passing through the two logic points, with a steep and symmetrically positioned central inflection part, surrounded by broad small-slope regions at the logic points, and with no measurable noise in the curve itself (unlike in the actual experimental data). The extensions of the curve indicate that the response could also be considered and measured somewhat beyond the logic points, if physically relevant. *The schematic* outlines the experimental system, "color-coded" to the plots. The *Red* and *Ox* labels refer to the redox states of the chromogen, TMB; DHA referes to dehydroascorbic acid — the product of irreversible oxidation of ascorbate (Asc).

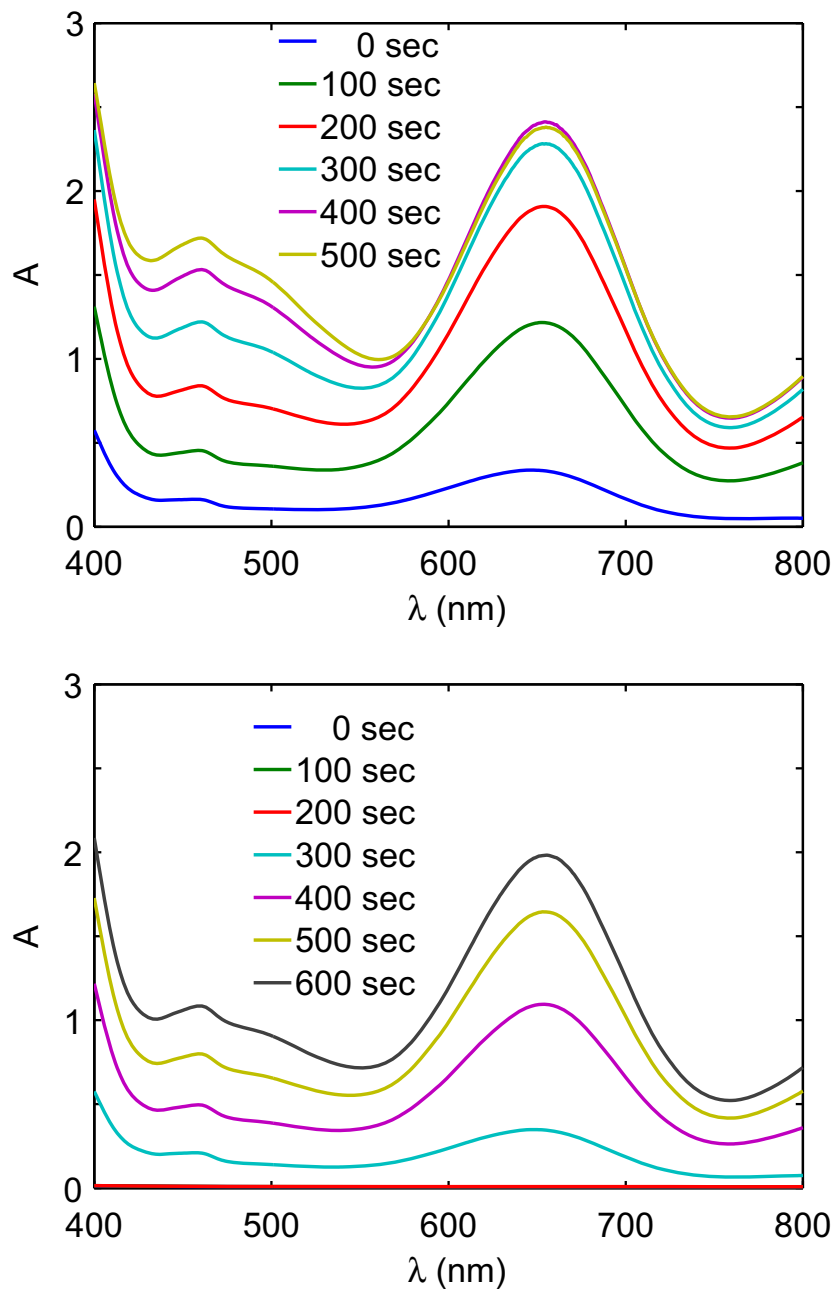

**Figure 2.** Absorbance, A, spectra at different reaction times. *Top panel:* without ascorbate in the solution. Note that due to high catalytic activity of the enzyme, without ascorbate there is a certain nonzero signal value already at the time at which it was experimentally practical to take the 0 sec measurement. *Bottom panel:* with 100 μM of ascorbate initially. Here the initial signal is very weak (the curves for 0 and 100 sec are obscured by the curve for 200 sec). Therefore, these data were taken for somewhat longer times.

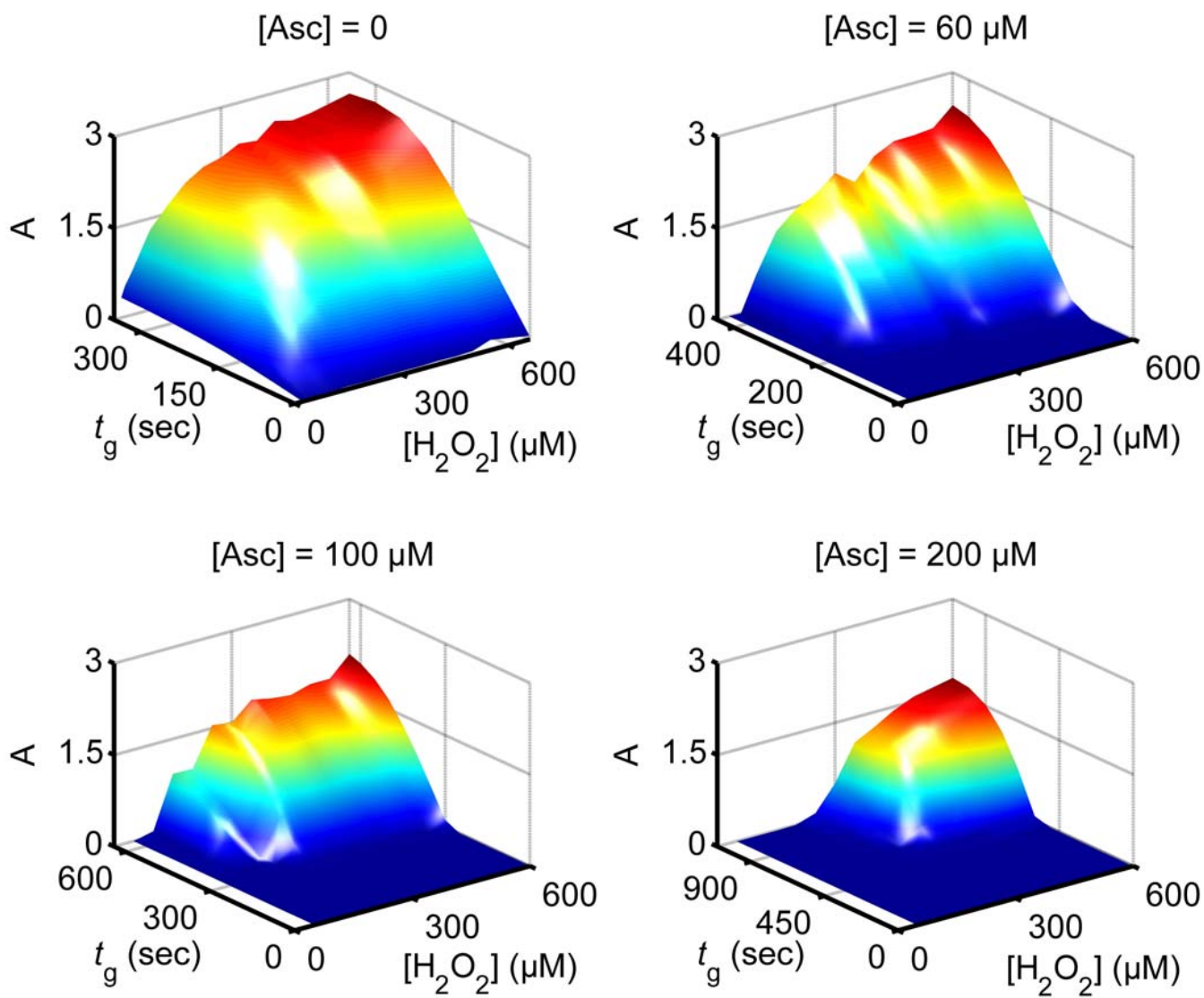

**Figure 3.** Experimental dependence of the concentration of the charge transfer species (the blue product), measured by the absorbance, A, on the initial concentration of $H_2O_2$, for varying reaction time, $t_g$, with different initial amounts of ascorbate, the concentration of which is shown above each plot.

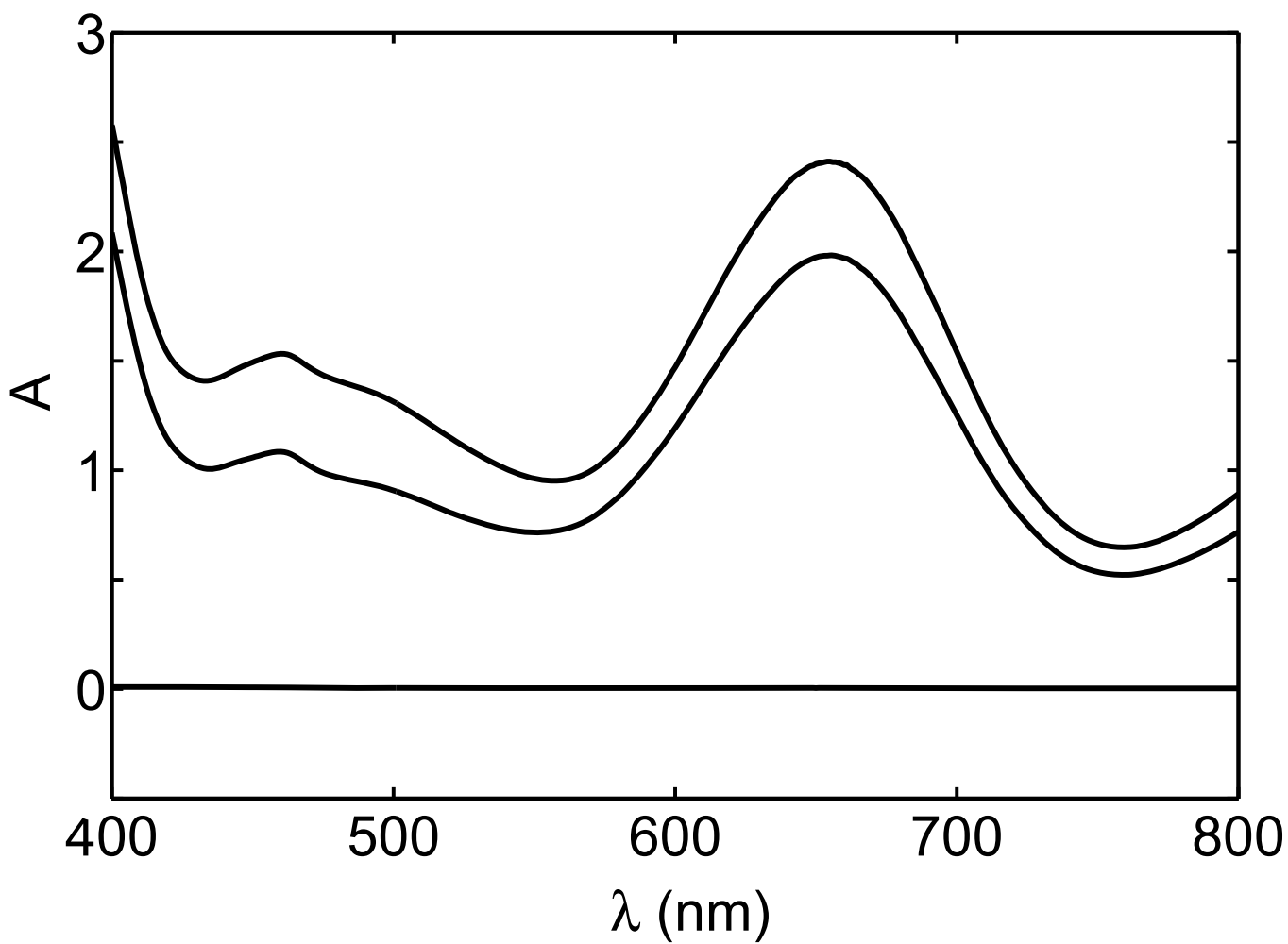

**Figure 4.** Selected spectral data. *Bottom:* Absorbance without addition of $H_2O_2$. The output signal is then zero at all the frequencies (here the curve is for data taken at 600 sec). *Middle:* These data were taken at 500 sec, for $[H_2O_2]$ at its initial "logic-1" value of 600 μM, without ascorbate present. *Top:* Data taken at 600 sec, with, initially, $H_2O_2$ at "logic-1" and 100μM of ascorbate added to the solution.

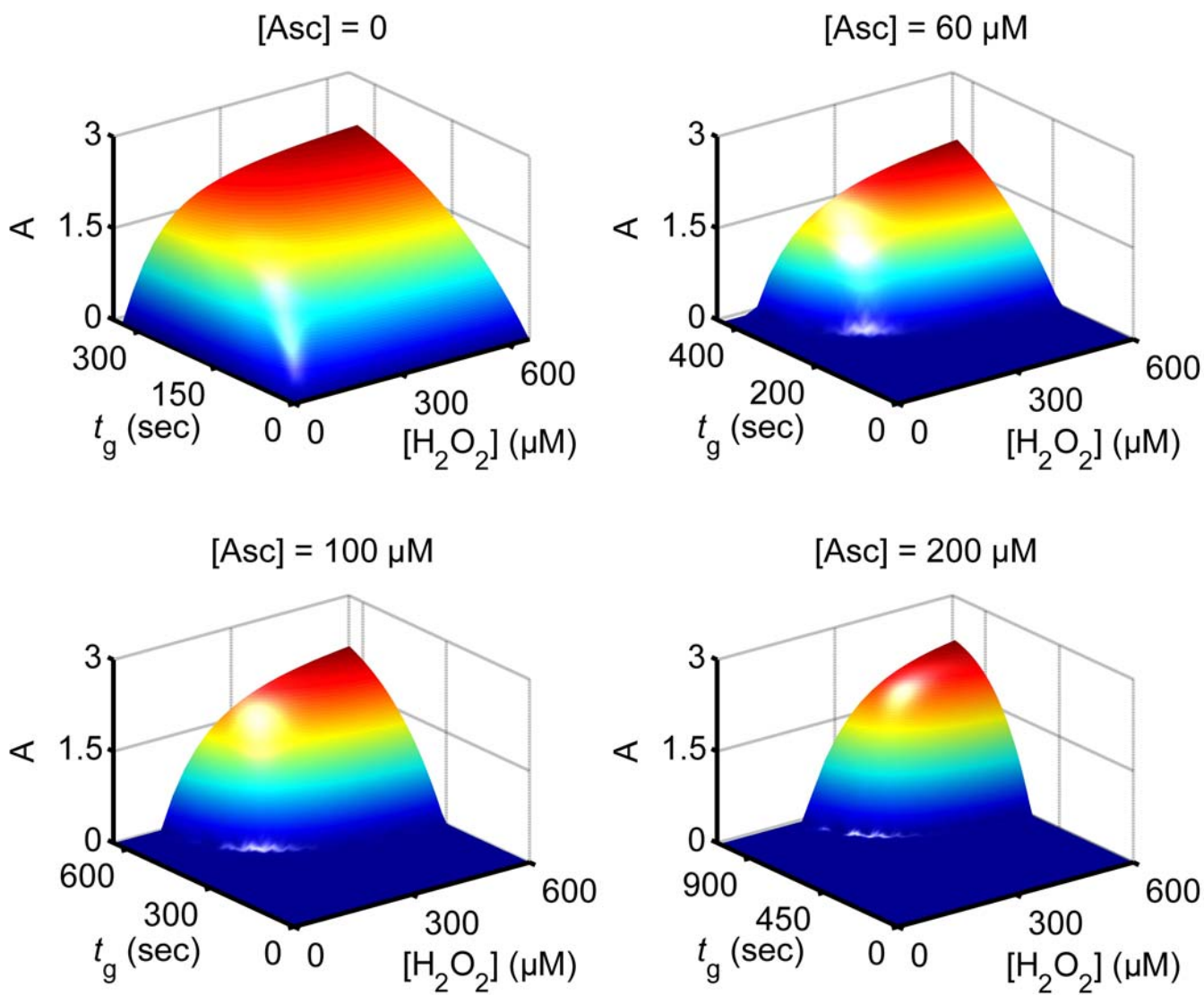

**Figure 5.** Fit of the data shown in Fig. 3 with the model rate equations, Eq. 5, corresponding to the processes identified in Eqs. 3-4. The fitted rate constant values are given in the text.

– 23 –

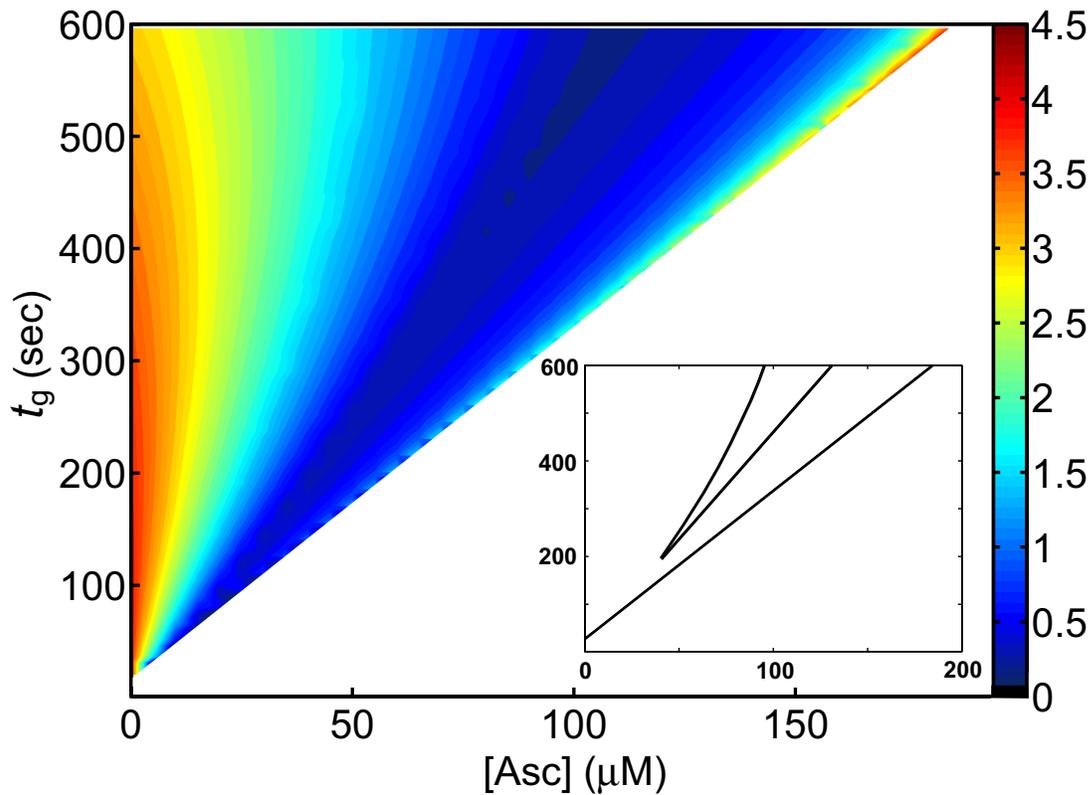

**Figure 6.** Values of the noise scaling factor, $\sigma_{out}^{max}/\sigma_{in}$, coded in the upper-left part of the plane of the ascorbate concentration and reaction (gate) time, according to the colors given in the vertical bar, for $\sigma_{in}=0.1\,(=10\%)$. The lower-right part of the plane is not color-coded: It corresponds to the regime of small output signal range (large intrinsic noise). The inset identifies the range of the optimal parameter values for filter operation at the assumed 10% input-noise level, delineated by the two connecting curves. The lower curve shows the boundary of the intrinsically noisy regime.